
\magnification=\magstep1
\overfullrule=0pt
\font\huge=cmr10 scaled \magstep2
\def\Eqdef{\,{\buildrel \rm def \over =}\,}
\font\smal=cmr7     
\def\z{{\cal Z}}  \def\la{\lambda} \def\r{{\bar{r}}}  \def\k{\bar{k}}
\def\L{\Lambda}  \def\J{{\cal J}}  \def\p{{\cal P}}
\def\eps{{\epsilon}}

\catcode`\@=11
\font\tenmsa=msam10
\font\sevenmsa=msam7
\font\fivemsa=msam5
\font\tenmsb=msbm10
\font\sevenmsb=msbm7
\font\fivemsb=msbm5
\newfam\msafam
\newfam\msbfam
\textfont\msafam=\tenmsa  \scriptfont\msafam=\sevenmsa
  \scriptscriptfont\msafam=\fivemsa
\textfont\msbfam=\tenmsb  \scriptfont\msbfam=\sevenmsb
  \scriptscriptfont\msbfam=\fivemsb
\def\hexnumber@#1{\ifcase#1 0\or1\or2\or3\or4\or5\or6\or7\or8\or9\or
	A\or B\or C\or D\or E\or F\fi }

\def\Bbb{\ifmmode\let\next\Bbb@\else
 \def\next{\errmessage{Use \string\Bbb\space only in math mode}}\fi\next}
\def\Bbb@#1{{\Bbb@@{#1}}}
\def\Bbb@@#1{\fam\msbfam#1}
\def\Z{{\Bbb Z}}

{\nopagenumbers
\rightline{November, 1995}\bigskip\bigskip
\centerline{{\bf \huge  The Level Two and Three Modular Invariants of SU(n)}}
\bigskip\bigskip\centerline{{Terry Gannon}\footnote{$^{\dag}$}{{\smal
Permanent address as of Sept 1996: Math Dept, York University, North York,
Canada M3J 1P3}}}
\medskip\centerline{{\it Max-Planck-Institut f\"ur Mathematik, Bonn,
D-53225}}
\bigskip \bigskip\centerline{{\bf Abstract}}\bigskip
In this paper we explicitly classify all modular invariant partition
functions for $A_r^{(1)}$ at level 2 and 3. Previously, these were known
only for level 1. The level 2 exceptionals exist at $r=9$, 15, and 27;
the level 3 exceptionals exist at $r=4$, 8, and 20. One of these is new,
but  the others were all anticipated by
the ``rank-level duality'' relating $A_r^{(1)}$ level $k$ and
$A_{k-1}^{(1)}$ level $r+1$. The main recent result which this paper rests on
is the classification of ``${\cal ADE}_7$-type invariants''.
\vfill\eject}

\centerline{{\bf 1. Introduction}}\medskip

The problem of classifying all
modular invariant partition functions corresponding to a given affine
algebra $X_r^{(1)}$ and level $k$, has been around for about a decade.
The problem is easy to state. The level $k$ highest weights $\la$ of
$X_r^{(1)}$ form a finite set $P_+^k(X_r^{(1)})$; the problem is to
find all sesquilinear combinations
$$\z=\sum_{\la,\mu\in P_+^k(X_r^{(1)})} M_{\la,\mu}\,\chi_\la\,\chi_\mu^*
\eqno(1)$$
of the characters $\chi_\la$, which satisfy three properties:\smallskip

\item{$\bullet$} $\z$ is modular invariant (i.e.\ its coefficient matrix $M$
commutes with the matrices $S$ and $T$ defined in eqs.(3) below);

\item{$\bullet$} $M_{\la,\mu}\in\Z_{\ge}\Eqdef\{0,1,2,\ldots\}$ for all $\la,
\mu\in P_+^k(X_r^{(1)})$;

\item{$\bullet$} $M_{k\L_0,k\L_0}=1$.\smallskip

\noindent These functions $\z$ or matrices $M$ are called {\it physical
invariants}.

In spite of considerable effort, for few $X_r^{(1)}$ and $k$ do we have
a complete classification. The original result is the A-D-E classification
[1] for $A_1^{(1)}$, $\forall k$. Also, $A_2^{(1)}$ for all $k$ is
known [2], as is $k=1$ for all (simple) $X_r^{(1)}$ [3,4]. In this paper we add
two more results: $k=2,3$ for all $A_r^{(1)}$.

Nevertheless, there has been considerable progress, on a more abstract
level, towards solving this problem. In particular, in [5] we find for
$A_r^{(1)}$ at all levels $k$, all physical invariants satisfying in
addition the condition
$$M_{\la,k\L_0}\ne 0\ {\rm or}\ M_{k\L_0,\la}\ne 0\quad\Longrightarrow
\quad \la=J'(k\L_0)\eqno(2)$$
for some simple current $J'$ (the simple currents for $A_r^{(1)}$ are
simply the rotation symmetries of its extended Coxeter-Dynkin diagram).
These physical invariants are called ${\cal ADE}_7${\it -type invariants}.
Almost every physical invariant is expected to satisfy eq.(2).

This paper is essentially two corollaries to [5]. The main purpose here is
simply to
illustrate the value of the ${\cal ADE}_7$ classification, and to help
clarify the final step in the physical invariant classification: the
determination of anomolous ``$\rho$-couplings''. Another immediate
consequence of [5] would be physical invariant classifications for $A_r^{(1)}$
for several ``small'' pairs $(r,k)$ (though admittedly, the value
of those results is not so clear).

In the following section we review the basic tools we will use, and list
all of the level $k\le 3$ physical invariants for $A_r^{(1)}$. In section
3 we give the completeness proof for $k=2$, and in section 4 we give it for
$k=3$.

\bigskip\centerline{{\bf 2. Review}}\medskip

 The level $k$ weights $\la$ form the set
$$P_+^{r,k}=P_+^k(A_r^{(1)})=\{(\la_0,\la_1,\ldots,\la_r)\,|\,\la_i\in
\Z_{\ge},\ \sum_i\la_i=k\,\}\ .$$
We will also write these as $\la=\sum\la_i\L_i$. For convenience put
$\L^i=(k-1)\L_0+\L_i$.
Define $\rho=(1,1,\ldots,1)$ and put $\r=r+1$, $\k=k+\r$. The affine
characters $\chi_\la$ at fixed level $k$ define a natural unitary
representation of $SL_2(\Z)$ [6]:
$$\eqalignno{\left(\matrix{0&-1\cr 1 &0}\right)\circ \chi_\la=&\,
\sum_{\mu\in P_+^{r,k}} S_{\la,\mu}\,\chi_\mu\ &(3a)\cr
\left(\matrix{1&1\cr 0 &1}\right)\circ \chi_\la=&\,
\sum_{\mu\in P_+^{r,k}} T_{\la,\mu}\,\chi_\mu\ .&(3b)\cr}$$
Then $\z$ in (1) is modular invariant iff
$$\eqalignno{M\,T=&\,T\,M&(4a)\cr M\,S=&\,S\,M\ .&(4b)}$$
(4a) is equivalent to the selection rule
$$M_{\la,\mu}\ne 0\quad \Longrightarrow\quad (\overline{\la+\rho}\,|\,
\overline{\la+\rho})\equiv (\overline{\mu+\rho}\,|\,\overline{\mu+\rho})
\qquad ({\rm mod}\ 2\k)\ ,\eqno(4c)$$
where $(-|-)$ denotes the familiar invariant form of $A_r$ (normalized so
that roots have norm 2), and $\bar{\la}=(\la_1,\ldots,\la_r)$. Eq.(4b) is
more difficult to interpret, though explicit expressions for $S_{\la,\mu}$
exist [6]. $S$ is unitary and symmetric.

Let $J$ and $C$ denote the following permutations of $P_+^{r,k}$:
$$\eqalignno{J(\la_0,\la_1,\ldots,\la_r)=&\,(\la_r,\la_0,\la_1,\ldots,
\la_{r-1})&\cr
C(\la_0,\la_1,\ldots,\la_r)=&\,(\la_0,\la_r,\la_{r-1},\ldots,\la_{1})\ .&\cr}$$
They are called simple currents and conjugations, respectively. Let
$\J_d$ denote the group generated by $J^d$. Write $[\la]$ for the orbit of
$\la$ with respect to
 both $C$ and $J$, and write $[\la]_d$ for the orbit with respect
to $\J_d$. $C$ defines a physical invariant in the obvious way, also denoted
by $C$. In a more subtle way [7], so does $J^d$. First define
$$t(\la)=\sum_{j=1}^rj\la_j\ ,$$
and put $k'=k$ unless both $\r$ and $k$ are odd, in which case put
$k'=\k$. Then for any divisor $d$ of $\r$ for which $k'd$ is even, we get
a physical invariant $I[\J_d]$ given by
$$I[\J_d]_{\la,\mu}=\sum_{j=1}^{\r/d}\delta^{\r/d}(t(\la)+d\,jk'/2)\,
\delta_{\mu,J^{jd}\la}\eqno(5)$$
where $\delta^y(x)=1$ or 0 depending respectively on whether or not $x/y\in
\Z$. Any physical invariant which cannot be expressed as the matrix product
$C^a\cdot I(\J_d)$ for some $a,d$, is called {\it exceptional}.

The level 2 exceptionals ${\cal E}^{(r,2)}$ are
$$\eqalignno{{\cal E}^{(9,2)}=&\,\sum_{i=0}^9|\chi_{J^i\L^0}+\chi_{J^i(\L_3+
\L_7)}|^2+\sum_{i=0}^4|\chi_{J^i\L^3}+\chi_{J^i(\L_5+\L_8)}|^2&(6a)\cr
{\cal E}^{(15,2)}=&\,\sum_{i=0}^7\bigl(|\langle\chi_{J^i\L^0}\rangle_8|^2+
|\langle\chi_{J^i\L^4}\rangle_8|^2+|\langle\chi_{J^i\L^6}\rangle_8|^2&\cr
&+|\chi_{J^i\L^8}|^2+\langle\chi_{J^i(\L_3+\L_5)}\rangle_8\,
\chi_{J^i\L^8}^*+\chi_{J^i\L^8}\,\langle\chi_{J^i(\L_3+\L_5)}\rangle_8^*\bigr)
&(6b)\cr
{\cal E}^{(27,2)}=&\,\sum_{i=0}^{13}\bigl(|\langle\chi_{J^i\L^0}\rangle_{14}+
\langle\chi_{J^i(\L_5+\L_{23})}\rangle_{14}|^2
+|\langle\chi_{J^i(\L_3+\L_{25})}\rangle_{14}+\langle\chi_{J^i(\L_6+\L_{22})
}\rangle_{14}|^2\bigr)&(6c)\cr}$$
together with the matrix products $C\cdot {\cal E}^{(9,2)}$, $C\cdot {\cal
E}^{(15,2)}$, ${1\over 2}I[\J_4]\cdot{\cal E}^{(15,2)}$, and $C\cdot
{\cal E}^{(29,2)}$. In these equations we use the short-hand
$$\langle\chi_\la\rangle_d\Eqdef\sum_{j=1}^{\r/d}\chi_{J^{jd}\la}\ .$$
${\cal E}^{(9,2)}$ first appeared in [8], which also anticipated the other
two -- although to this author's knowledge neither ${\cal E}^{(15,2)}$ nor
${\cal E}^{(29,2)}$ have appeared explicitly in the literature before (however
[9] found the projection ${1\over 2}I[\J_4]\cdot{\cal E}^{(15,2)}$). The level
 3 exceptionals are
$$\eqalignno{{\cal E}^{(4,3)}=&\,\sum_{i=0}^4\bigl(|\chi_{J^i\L^0}+
\chi_{J^i(\L_0+\L_2+\L_3)}|^2+|\chi_{J^i(2\L_1+\L_3)}+\chi_{J^i
(\L_2+2\L_4)}|^2\bigr)&(7a)\cr
{\cal E}^{(8,3)}=&\,\sum_{i=0}^2\bigl(|\langle\chi_{J^i\L^0}\rangle_3|^2+
|\langle\chi_{J^i\L^3}\rangle_3|^2+|\langle\chi_{J^i\L^6}\rangle_3|^2&\cr
&+|\langle\chi_{J^i(\L_0+\L_1+\L_5)}\rangle_3|^2+|\langle\chi_{J^i
(\L_0+\L_2+\L_4)}\rangle_3|^2+2|\chi_{J^i(\L_0+\L_3+\L_6)}|^2&\cr
&+\langle\chi_{J^i(\L_2+\L_3+\L_4)}\rangle_3\,
\chi_{J^i(\L_0+\L_3+\L_6)}^*+\chi_{J^i(\L_0+\L_3+\L_6)}\langle\,
\chi_{J^i(\L_2+\L_3+\L_4)}\rangle_3^*\bigr)&(7b)\cr
{\cal E}^{(8,3)}{}'=&\,\sum_{i=0}^{2}\bigl(|\langle\chi_{J^i\L^0}\rangle_{3}+
\langle\chi_{J^i(\L_0+\L_4+\L_{5})}\rangle_{3}|^2+2|\langle
\chi_{J^i(\L_0+\L_2+\L_{7})}\rangle_{3}|^2\bigr)\ ,&(7c)\cr
{\cal E}^{(8,3)}{}''=&\,|\langle\chi_{\L^0}\rangle_{3}+
\langle\chi_{\L_0+\L_4+\L_{5}}\rangle_{3}|^2+|\langle\chi_{\L_0+\L_2+
\L_{7}}\rangle_{3}+\langle\chi_{\L_0+\L_2+\L_4}\rangle_3|^2& \cr
&+\langle\chi_{\L_1+\L_3+\L_5}\rangle_3\,
\bigl(\sum_{i=1}^2\langle\chi_{J^i\L^0}
\rangle_3 +\langle\chi_{J^i(\L_0+\L_4+\L_5)}\rangle_3\bigr)^*&\cr&+
\bigl(\sum_{i=1}^2\langle\chi_{J^i\L^0}
\rangle_3 +\langle\chi_{J^i(\L_0+\L_4+\L_5)}\rangle_3\bigr)\,
\langle\chi_{\L_1+\L_3+\L_5}\rangle_3^*&(7d)\cr
{\cal E}^{(20,3)}=&\sum_{i=0}^{6}\bigl(|\langle\chi_{J^i\L^0}\rangle_{7}+
\langle\chi_{J^i(\L_0+\L_4+\L_{17})}\rangle_{7}+\langle\chi_{J^i(\L_0+\L_6
+\L_{15})}\rangle_{7}+\langle\chi_{J^i(\L_0+\L_{10}+\L_{11})}\rangle_{7}|^2
&(7e)\cr
&+|\langle\chi_{J^i(\L_1+\L_8+\L_{12})}\rangle_{7}+
\langle\chi_{J^i(\L_9+\L_{13}+\L_{20})}\rangle_{7}+
\langle\chi_{J^i(2\L_2+\L_{17})}\rangle_{7}+\langle
\chi_{J^i(\L_4+2\L_{19})}\rangle_{7}|^2&\cr}$$
together with their conjugates. To this author's knowledge, only ${\cal E}^{(
8,3)}$ has explicitly appeared in the literature (in [9]), though certainly
${\cal E}^{
(4,3)}$, ${\cal E}^{(8,3)}{}'$ and ${\cal E}^{(20,3)}$ have been anticipated.
The exceptional ${\cal E}^{(8,3)}{}''$ appears to be completely new.

The main result of this paper is that these exhaust all physical invariants
for $A_r^{(1)}$ at levels 2,3. At level 1, there are no exceptionals:
all physical invariants are given by eq.(5). Note the strong resemblance of
the exceptionals in eqs.(6),(7) (except ${\cal E}^{(8,3)}{}''$) to the
exceptionals of $A_1^{(1)}$ [1]
and $A_2^{(1)}$ [2], respectively. This is not a coincidence, and is a
consequence of  a duality [8,10,5] between $A_r^{(1)}$ level $k$, and
$A_{k-1}^{(1)}$
level $r+1$. In particular, choose any $\la\in P_+^{r,k}$. Construct its
Young diagram, and reflect it through the diagonal (i.e.\ take its transpose).
Deleting all columns (if any) of length $k$, this will be the Young diagram
of some weight in $P_+^{k-1,r+1}$ which we will denote by $T(\la)$. To avoid
confusion
{\it we will generally put tilde's over the quantities of $A_{k-1}^{(1)}$
level $r+1$}. Now define a map $T':P_+^{r,k}\rightarrow P_+^{k-1,r+1}$ by
$T'(\la)=\tilde{J}^{-(t(\la)-\{t(\la)\})/\r}T(\la)$, where $\{t(\la)\}$
denotes the smallest nonnegative integer such that $t(\la)\equiv \{t(\la)\}$
(mod $\r$). Then
$$S_{\la,\mu}=\sqrt{{k\over\r}}\exp[{2\pi i\over \r k}\{t(\la)\}\{t(\mu)\}]\,
\tilde{S}^*_{T'(\la),T'(\mu)}\eqno(8)$$
with a similar expression relating $T_{\la,\mu}$ and $\tilde{T}_{T'(\la),
T'(\mu)}$. From (8a) we find that $T'(J\la)\in \tilde{\J}_1T'(\la)$. With
the exception of ${\cal E}^{(8,3)}{}''$, the map $T'$ connects the
exceptionals of $A_1^{(1)}$ and $A_2^{(1)}$ with those in eqs.(6),(7) above.

We conclude this section by reviewing the basic lemmas. A well-known result
is [6]
$$S_{\L^0,\la}\ge S_{\L^0,\L^0}>0\ ,\eqno(9a)$$
with equality in (9a) iff $\la\in[\L^0]$. Two useful identities are
$$\eqalignno{S_{J^a\la,J^b\mu}=&\,\exp[2\pi i\,(b\,t(\la)+a\,t(\mu)+kab)/\r]\,
S_{\la,\mu}&(9b)\cr
{(\overline{J^a\la+\rho}\,|\,\overline{J^a\la+\rho}) \over 2\k}\equiv&\,
{-2a\,t(\la)+ka\,(\r-a)\over 2\r}+{(\overline{\la+\rho}\,|\,\overline{\la+\rho}
)\over 2\k}\ .&(9c)\cr}$$

For the remainder of this section let $M$ be any physical invariant. Useful
definitions are
$$\eqalignno{\p_L=&\,\{\la\in P_+^{r,k}\,|\,\exists \mu\in P_+^{r,k}\ {\rm
such}\ {\rm that}\ M_{\la,\mu}\ne 0\,\}&\cr
s_L(\la)=&\,\sum_{\mu\in P_+^{r,k}} S_{\la,\mu} M_{\mu,\L^0}&\cr}$$
and define $\p_R$ and $s_R(\mu)$ analogously. Our first result
comes from [5,11].

\smallskip\noindent{{\bf Cyclotomy Lemma}} \quad (a) For each $\la\in
P_+^{r,k}$, $s_L(\la)\ge 0$. Also, $s_L(\la)>0$ iff $\la\in \p_L$.

\noindent{(b)} If $M_{J^a\L^0,J^b\L^0}\ne 0$, then $M_{J^a\la,J^b\mu}=M_{\la,
\mu}$ for all $\la,\mu\in P_+^{r,k}$, and $M_{\la,\mu}\ne 0$ only if
$a\,t(\la)\equiv b\,t(\mu)$ (mod $\r$).\smallskip

We may decompose $M$ into a direct sum of submatrices $M^{(i)}$, and apply
Perron-Frobenius theory [12] to each submatrix. Let $M^{(0)}$ be the
submatrix containing the index $\L^0$, and let $e(M^{(i)})$ denote the
largest real eigenvalue of $M^{(i)}$.

\smallskip\noindent{{\bf Perron-Frobenius Lemma} [11]} \quad $e(M^{(i)})\le
e(M^{(0)})$ for all $i$. If $(M^{(0)})^2=e\,M^{(0)}$ for some number $e$,
then $e(M^{(i)})=e$ for each nonzero $M^{(i)}$.\smallskip

A final very important result is the Galois symmetry [13] obeyed by $S$.
Choose any $\la\in P_+^{r,k}$, and any integer $\ell$ coprime to $\r\k$.
Then there will exist a Weyl group element $w$ of $A_r$, a root lattice
element $\alpha$ of $A_r$, and a weight in $P_+^{r,k}$ which we will denote
by $(\ell\la)_+$, for which
$$\eqalignno{\ell\,\,\overline{\la+\rho}=&\,w(\overline{(\ell\la)_+})+
\k\alpha &(10a)\cr
\eps_\ell(\la)\,S_{(\ell\la)_+,\mu}=&\,\eps_\ell(\mu)\,S_{\la,(\ell\mu)_+}
\qquad\forall \la,\mu\in P_+^{r,k}\ ,&(10b)\cr}$$
where $\eps_\ell(\la)={\rm det}\,w\in\{\pm 1\}$. Eq.(10b) also equals the
value of the Galois automorphism corresponding to $\ell$, applied to
$S_{\la,\mu}$ (upto in irrelevant sign independent of $\la$ and $\mu$).
This is important because rank-level duality and (9a) then imply that
$$\eqalignno{\eps_\ell(\la)\,\eps_\ell(\L^0)=&\,\tilde{\eps}_\ell(T'(\la))\,
\tilde{\eps}_\ell(\tilde{\L}^0)&(10c)\cr
T'((\ell\la)_+)\in &\,\tilde{\J}_1(\ell T'(\la))_+\ .&(10d)\cr}$$
In particular, (10c) follows by applying the Galois automorphism to
$S_{\la,\L^0}/S_{\L^0,\L^0}=\tilde{S}_{T'(\la),\tilde{\L}^0}/\tilde{S}_{
\tilde{\L}^0,\tilde{\L}^0}$. Eq.(10d) follows because rank-level duality $T'$
is an exact (ignoring the irrelevant $\sqrt{k/\r}$ factor) bijection for
weights $\mu\in P_+^{r,k}$ with $t(\mu)\equiv 0$ (mod $\r$): that implies
$$\tilde{S}_{T'((\ell\la)_+),\tilde{\mu}}=\tilde{S}_{\ell (T'(\la))_+,
\tilde{\mu}}$$
for all $\tilde{\mu}\in P_+^{k-1,r+1}$ with $\tilde{t}(\mu)\equiv 0$ (mod $k$).
{}From this equation, (10d) immediately follows.

\smallskip\noindent{{\bf Galois Lemma} [4,14]}\quad (a) $M_{\la,\mu}\ne 0$
only if $\eps_\ell(\la)=\eps_\ell(\mu)$ for all $\ell$ coprime to $\r\k$.

\noindent{(b)} $M_{\la,\mu}=M_{(\ell\la)_+,(\ell\mu)_+}$ for all $\la,\mu
\in P_+^{r,k}$, and all $\ell$ coprime to $\r\k$.\smallskip

We are most interested in applying Galois (a), together with Cyclotomy (a)
and eq.(4c), to find the possibilities $\la$ with either $M_{\la,\L^0}$
or $M_{\L^0,\la}$ nonzero. What we will find is that, for all but finitely
many pairs $(r,2)$, $(r,3)$, eq.(2) will necessarily be satisfied. In [5],
all such physical invariants were classified; for $k\le 3$ we found that
they are either of the form $C^a\cdot I[\J_d]$, or equal $C^a\cdot{\cal E}^{
(15,2)}$, $C^a\cdot{\cal E}^{(8,3)}$, or ${1\over 2}I[\J_4]\cdot{\cal E}^{
(15,2)}$.\bigskip

\centerline{{\bf 3. The level 2 physical invariants of $A_r^{(1)}$}}\medskip

Throughout this section let $M$ denote any physical invariant of $A_r^{(1)}$
level 2, not satisfying eq.(2). The condition $\eps_\ell(\la)=\eps_\ell(\L^0)$
in Galois (a) was explicitly solved for $A_1^{(1)}$ in [11]. The result is
that it forces $\la\in[\L^0]_1$, unless $\k=6$, 10, 12 or 30. Thus by (10c),
it suffices to consider $r=3$, 7, 9 or 27. $r=3$ and $r=7$ are handled
directly by (4c).

\smallskip
\noindent{{\it $A_9^{(1)}$ level 2:}}\quad  By Galois (a), the only
possibilities $\la$ with $M_{\la,\L^0}$ or $M_{\L^0,\la}$ nonzero are
$\la\in[\L^0]\cup[\L^4]$. Eq.(4c) now forces $\la=\L^0$ or $\la=\L_3+\L_7\Eqdef
\la'$. Now compute $s_L(\L^1)$ -- this is trivial using (8). Then Cyclotomy
(a) tells us
$$\sin(\pi/6)-M_{\la',\L^0}\sin(\pi/6)\ge 0\ .\eqno(11a)$$
Thus $M_{\la',\L^0}=0$ or 1. Similarly for $M_{\L^0,\la'}$. But (4b)
evaluated at $(\L^0,\L^0)$ forces $M_{\la',\L^0}=M_{\L^0,\la'}=1$.

Next, consider the possible $\la$ with $M_{J\L^0,\la}\ne 0$. Again from
Galois (a) and (4c), we find that the only possibilities are $\la\in\{
J^{\pm 1}\L^0,J^{\pm 1}\la'\}$. Now if $M_{J\L^0,J^i\L^0}=0$ for all $i$,
then (4b) evaluated at $(J\L^0,\L^0)$ would give
$$(M_{J\L^0,J\la'}+M_{J\L^0,J^{-1}\la'}-1)\,S_{\la',\L^0}=S_{\L^0,\L^0}\ ,
\eqno(11b)$$
which contradicts (9a). Thus, multiplying $M$ if necessary by $C$, we may
assume $M_{J\L^0,J\L^0}\ne 0$.

Note that $M_{\la',\la'}=1$ by Galois (b) with $\ell=7$ (again this is easiest
to see using rank-level duality).
Finally, consider the possible  $\la$ with $M_{\L^3,\la}\ne 0$. By Galois (a),
$\la\in[\L^3]$, so by Cyclotomy (b), $\la\in\{\L^3,J^5\L^3\}$. Putting
$(\L^3,\L^1)$ in (4b) forces $M_{\L^3,J^5\L^3}=M_{\L^3,\L^3}$, and now
Perron-Frobenius forces them to equal 1.

All other entries of $M$ are fixed by Cyclotomy (b), and we find $M={\cal
E}^{(9,2)}$.

\smallskip\noindent{{\it $A_{27}^{(1)}$ level 2:}}\quad As before, Galois (a)
and (4c) tell us $M_{\L^0,\la}\ne 0$ requires $\la\in\{[\L^0]_{14},[\la_5+
\la_{23}]_{14}\}$. Put $\la'=\la_5+\la_{23}$. Assume first that $M_{\L^0,
J^{14}\L^0}=0$. Write $m=M_{\L^0,\la'}$, $m'=M_{\L^0,J^{14}\la'}$. Then
$s_L(\L^{j-1})\ge 0$ requires
$$\sin(\pi j/30)+m\sin(11\pi j/30)+m'\sin(19\pi j/30)\ge 0\ .\eqno(12)$$
We get $m=m'=0$ by taking $j=2,4,15$ in (12). Similarly for $M_{\la,\L^0}
\ne 0$. Hence  $M$ will obey eq.(2), contrary to hypothesis.

Thus $M_{\L^0,J^{14}\L^0}=M_{J^{14}\L^0,\L^0}=1$. Hence by Cyclotomy (b),
$M_{\L^0,\la'}=M_{\L^0,J^{14}\la'}=m$. $s_L(\la^{15})\ge 0$ forces $m=1$.

As before (by conjugating $M$ if necessary),  we may assume $M_{J\L^0,J\L^0}
\ne 0$. $M_{\la',\la'}=1$ is forced by Galois (b) with $\ell=11$. All other
entries of $M$ are now found by Cyclotomy (b) and Galois (b) ($\ell=13$).
We obtain ${\cal E}^{(27,2)}$.\bigskip

\centerline{{\bf 4. The level 3 physical invariants of $A_r^{(1)}$}}\medskip

Throughout this section let $M$ denote any physical invariant of
$A_r^{(1)}$ level 3 which does not satisfy eq.(2). The condition $\eps_\ell
(\la)=\eps_\ell(\L^0)$ was also solved for $A_2^{(1)}$ in [11], though in
places the proof used (4c). Fortunately we are saved the hassle of verifying
that the argument in [11] also works if one replaces (4c) with its rank-level
dual, by a remarkable coincidence discovered (I believe) by Ph.\ Ruelle:
The Galois condition $\eps_\ell(\la)=\eps_\ell(\mu)$ for $A_2^{(1)}$ also
appears naturally in the analysis of Jacobians of Fermat curves! In particular,
Thm.\ 0.3 of [15] solves this condition for all but 31 levels $k$ (the
highest exception being $k=177$).

For now let us avoid these 31 anomolous $k$. Then we learn from [15] that
for $\k$ odd, only $\la\in[\L^0]$ satisfies Galois (a) with $\mu=\L^0$.
When $\k\equiv 0$ (mod 4), we get $\la\in[\L^0]\cup[\L_0+2\L_s]
\cup [\L_0+\L_s+\L_{2s}]$, and when $\k\equiv 2$ (mod 4), we get
$\la\in[\L^0]\cup[\L_0+2\L_s]\cup[\L_0+\L_s+\L_{2s}]\cup[\L_0+2\L_{(r-2)/4}]$,
where we put $s=r/2$.

Now apply (4c) to $\la\in[\L_0+2\L_s]$ and $\mu=\L^0$. Multiplying (4c) by
$2\r$ and using (9c), eq.(4c) implies $(\r^2-\r+1)/2-2/\k\in\Z$, which can
never hold. The other possibilities for $\la$ can be analysed similarly,
as can the 31 anomolous levels. What we find is that (4c) and Galois (a)
require any $\la$ satisfying $M_{\L^0,\la}\ne 0$ or $M_{\la,\L^0}\ne 0$ to
be:

\item{(i)} for $\k\not\equiv 0$ (mod 4): \quad $\la\in[\L^0]$;

\item{(ii)} for $\k\equiv 0$ (mod 4), $\k\ne 24,60$: \quad
$\la\in[\L^0]_d\cup[\la^s]_d$,
where $d$ is the smallest positive solution to $k'd^2\equiv 0$ (mod $2\r$);

\item{(iii)} $\k=24$: \quad $\la\in [\L^0]_7\cup[\la^4]_7\cup[\la^6]_7\cup
[\la^{10}]_7$;

\item{(iv)} $\k=60$: \quad $\la\in[\L^0]_{19}\cup[\la^{10}]_{19}\cup
[\la^{18}]_{19}\cup[\la^{28}]_{19}$,

\noindent where we put $\la^i=\L_0+\L_i+\L_{\r-i}$ in (ii)-(iv), and where
$k'$ in (ii) is defined near eq.(5).

The arguments applying eq.(4b) and Cyclotomy (a) reduce by eq.(8) to the
$A_2^{(1)}$
calculations explicitly given in [11]. We will give here one example.
Suppose $M_{\L^0,\la}=0$ unless $\la\in[\L^0]_d\cup[\la^s]_d$, as in (ii).
Define $m=\sum_{\la\in[\L^0]}M_{\L^0,\la}$, and
$m'=\sum_{\la\in[\la^s]}M_{\L^0,\la}$. Then
by Cyclotomy (b), $m'\ge m$. Now $s_R(2\L_1+\L_{r-1})\ge 0$ then reduces to
[11]
$$0\le (m+m')\sin[{2\pi\over\k}]+(m-m')\sin[{8\pi\over
\k}]-(m+m')\sin[{10\pi\over \k}]\le (m+m')\{\sin[{2\pi
\over \k}]-\sin[{10\pi\over\k}]\}\ .$$
This forces $\k=8$ or $\k=12$.

\smallskip\noindent{{\it $A_4^{(1)}$ level 3:}}\quad Here $d=5,m=1$. $s_R(\L^1)
\ge 0$ forces $m'=1$, and (4b) evaluated at $(\L^0,\L^0)$ forces $M_{\L^0,
\la^s}=M_{\la^s,\L^0}$. Conjugating if necessary, the usual argument forces
$M_{J\L^0,J\L^0}=1$. $M_{\la^s,\la^s}=1$ follows from Galois (b) $(\ell=s+1)$.
We are done if we know the values of $M_{2\L_1+\L_3,\la}$. For this to be
nonzero, Cyclotomy (b) and Galois (a) says $\la$ must equal either
$\mu^1\Eqdef2\L_1+\L_3$ or $C\mu^1$. Now (4b) evaluated at $(\L^1,\mu^1)$
gives us $M_{\mu^1,\mu^1}=M_{\mu^1,C\mu^1}$, and Perron-Frobenius
forces $M_{\mu^1,\mu^1}=1$.

\smallskip\noindent{{\it $A_8^{(1)}$ level 3:}}\quad The argument here is also
analogous to the calculations given in [11]. The only difference is when we
try to prove $M_{J\L^0,J\L^0}\ne 0$. Galois (a) and eq.(4c) permit an
unexpected possibility for $M_{J\L^0,\la}\ne 0$: $\la\in[\mu^3]_3$
where $\mu^3=\L_1+\L_3+\L_5$. If $M_{J\L^0,\mu^3}=0$, then all proceeds as
before, but if that entry is nonzero we get $M={\cal E}^{(8,3)}{}''$ by using
the familiar arguments.

\smallskip\noindent{{\it $A_{20}^{(1)}$ level 3:}}\quad Again the only
difference here arises in the proof that $M_{J\L^0,J\L^0}\ne 0$. If
$M_{J\L^0,J^i\L^0}=0$ for all $i$, then we get from (4b) evaluated at $(J\L^0,
\L^0)$ the equality
$${m_4-3\over 3}S_{\L^0,\la^4}+{m_6-3\over 3}S_{\L^0,\la^6}+{m_{10}-3\over 3}
S_{\L^0,\la^{10}}=S_{\L^0,\L^0}\ ,$$
where $m_i$ are non-negative integer multiples of 3 defined in the obvious
way. Since $S_{\L^0,\la^i}/S_{\L^0,\L^0}\approx 81.2,$ 137.9, 57.7 for
$i=4$, 6, 10, respectively, we see this requires $m_4=m_{10}=6$ and $m_6=0$.
This however is readily seen to violate Galois (b) (take $\ell=31$).

\smallskip\noindent{{\it $A_{56}^{(1)}$ level 3:}}\quad The identical
contradiction used in [11] works here.

\medskip\noindent{{\it Acknowledgements}} I thank Philippe Ruelle, who
brought Ref.[14] to my attention, and the MPIM for its hospitality.

\centerline{\bf References}\medskip

\item{[1]} A.\ Cappelli, C.\ Itzykson, and J.-B.\ Zuber, {\it Commun.\
Math.\ Phys.}\ {\bf 113} (1987), 1-26.

\item{[2]} T.\ Gannon, {\it Commun.\ Math.\ Phys.}\ {\bf 161} (1994),
233-264.

\item{[3]} P.\ Degiovanni, {\it Commun.\ Math.\ Phys.}\ {\bf 127} (1990),
71-99.

\item{[4]} T.\ Gannon, {\it Nucl.\ Phys.}\ {\bf B396} (1993), 708-736.

\item{[5]} T.\ Gannon, ``Kac-Peterson, Perron-Frobenius, and the
classification of conformal field theories'' (q-alg/9510026).

\item{[6]} V.\ G.\ Kac, {\it Infinite Dimensional Lie Algebras}, 3rd
edition, Cambridge University Press, Cambridge, 1990.

\item{[7]} A.\ N.\ Schellekens and S.\ Yankielowicz, {\it Phys.\ Lett.}\
{\bf B227} (1989), 387-391.

\item{[8]} M.\ A.\ Walton, {\it Nucl.\ Phys.}\ {\bf B322} (1989), 775-790.

\item{[9]} A.\ Font, {\it Mod.\ Phys.\ Lett.}\ {\bf A6} (1991), 3265-3272.

\item{[10]} D.\ Altschuler, M.\ Bauer and C.\ Itzykson, {\it Commun.\ Math.\
Phys.}\ {\bf 161} (1994), 233-264.

\item{[11]} T.\ Gannon, ``The classification of SU(3) modular invariants
revisited'', to appear in {\it Annales de l'I.H.P., Phys.\ Th\'eor.}\
(hep-th/9404185).

\item{[12]} F.\ R.\ Gantmacher, {\it The Theory of Matrices}, Chelsea
Publishing Co, New York, 1990.

\item{[13]} A.\ Coste and T.\ Gannon, {\it Phys.\ Lett.}\ {\bf B323} (1994),
316-321.

\item{[14]} Ph.\ Ruelle, E.\ Thiran and J.\ Weyers, {\it Nucl.\ Phys.}\
{\bf B402} (1993), 393-708.

\item{[15]} N.\ Aoki, {\it Amer.\ J.\ Math.}\ {\bf 113} (1991), 779-833.

\end